\documentclass[<review>]{elsarticle}
\usepackage[utf8]{inputenc}

\usepackage{color}
\usepackage{tabularx}
\usepackage{graphicx}
\usepackage{rotating}
\usepackage{enumerate}
\usepackage{float}
\usepackage{amsmath}
\usepackage{pstricks,pst-grad}
\usepackage{makeidx}
\usepackage{amsfonts}
\usepackage{multicol}
\usepackage{subcaption}
\usepackage{cite}
\usepackage{natbib}

\begin{document}
\begin{frontmatter}
\title{Fast Automatic Detection of Geological Boundaries from Multivariate Log Data Using Recurrence}
\author[1,2]{Ayham Zaitouny\corref{cor1}%
 \fnref{fn1}}
\ead{ayham.zaitouny@csiro.au}
\ead{ayham.zaitouny@uwa.edu.au}
\author[1,2,4]{Michael Small\fnref{fn2}}
\author[2]{June Hill\fnref{fn2}}
\author[3]{Irina Emelyanova\fnref{fn2}}
\author[3]{M. Ben Clennell\fnref{fn2}}
\address[1]{Mineral Resources, Commonwealth Scientific and Industrial Research Organisation, 26 Dick Perry Ave, Kensington WA 6151, Australia}
\address[2]{Complex Systems Group, Department of Mathematics and Statistics, University of Western Australia, 35 Stirling Highway, Crawley WA 6009, Australia}
\address[3]{Energy, Commonwealth Scientific and Industrial Research Organisation, 26 Dick Perry Ave, Kensington WA 6151, Australia}
\address[4]{Australian Centre for Transforming Maintenance through Data Science, University of Western Australia, Australia}
\cortext[cor1]{Corresponding author}
\fntext[fn1]{AZ initiated the idea, developed the method, performed the computation and wrote the manuscript.}
\fntext[fn2]{MS, JH, IE and BC initiated the idea, discussed the results, revised and improved the manuscript.}
\begin{abstract}
Manual interpretation of data collected from drill holes for mineral or oil and gas exploration is time-consuming and subjective.  Identification of geological boundaries and distinctive rock physical property domains is the first step of interpretation. We introduce a multivariate technique, that can identify geological boundaries from petrophysical or geochemical data. The method is based on time-series techniques that have been adapted to be applicable for detecting transitions in geological spatial data. This method allows for the use of multiple variables in detecting different lithological layers. Additionally, it reconstructs the phase space of a single drill-hole or well to be applicable for further investigations across other holes or wells. The computationally cheap method shows efficiency and accuracy in detecting boundaries between lithological layers, which we demonstrate using examples from mineral exploration boreholes and an offshore gas exploration well.
\end{abstract}
\begin{keyword}
Boundary detection, Drill holes, Well logs, Multivariate, Recurrence plot, Quadrant Scan
\end{keyword}
\end{frontmatter}

\section{Introduction}
Drill holes (for mineral exploration) or wells (for petroleum exploration) are the primary source for detailed information on the geology of the subsurface. Geological materials can be sampled and measured at regularly spaced intervals down the hole either by taking core samples directly during drilling or by inserting sensing tools into the hole to collect petrophysical logs either while drilling or after the completion of a well using a wireline system. The resultant spatial data (i.e. depth attributed data) is interpreted in terms of geological features. This process of interpretation has traditionally been done manually by geologists in the case of cores, and by petrophysicists in the case of downhole logs. However, human-based methods can result in inconsistent interpretations and the decision process is not transparent. It is also very time-consuming, both to identify the rock types and geological features, and also just to decide which sections require a new and different description from the previous sections. Geological interpretation can benefit from automation of many of the decision-making processes so that the results are reproducible and can be rapidly generated (see \citep{MLexample} as an example). This enables the interpreter to focus his skills on the meaning and implications of the interpretations, and on the most significant features for a particular application.
Geological rock units generally have sharp boundaries; e.g. boundaries between layers in a sedimentary sequence, or boundaries between intrusive rocks and host rocks. Across the boundary we see marked changes in chemical composition and/or in measured physical properties. The detection of these boundaries allows the geologist to define the extent of geological units, including rock units that may host valuable resources (e.g. oil and gas) and units that contain mineral deposits. Identification of boundaries is also a critical step in stratigraphic correlation of sedimentary deposits, whereby a change in the environment of deposition over time, or an event of erosion removing earlier layers (an unconformity), leads to one rock type being deposited on top of the previous rock type. This so-called \emph{litho}--stratigraphic correlation is used to determine sedimentary architecture within a basin. The physical properties changes across boundaries also produce the seismic reflections seen in geophysical surveys across the basin, so that well intersections can be tied into seismic images, enabling the strata to be mapped out in two or three dimensions. When samples containing fossils or radiometrically dateable materials can be obtained, then the geological interpretation can also make use of \emph{chrono}--stratigraphic correlation, based on matching the patterns of absolute ages, mainly using the emergence and extinction of microfossil species over hundreds of thousands or millions of years. We should point out that these changes in fossil populations may not necessarily reflected in any visible changes in the rock type or the physical properties, and indeed lithostratigraphic correlations are often discordant in time.

In this paper we present a method for the detection of geological boundaries (lithostratigraphy in the case of sediments, or a change in mineral characteristics in the case of igneous or metamorphic rocks) from drill hole or well log data. The method utilises non-linear time series change-point techniques to detect transitions between different geochemical or petrophysical domains along the length of the hole. The time series methods must be adapted to be suitable for 1 dimensional spatial data because some time-series procedures (such as embedding) are not directly applicable to spatial data.

Transitions or boundary detection methods are used to locate signal tipping points in various applications. In addition to detecting tipping points in time series for dynamical systems \citep{marr1980}, these detection algorithms can be used for non-temporal signals. For instance, detection techniques are used to identify object boundaries in image analysis \citep{canny1987}. They are also used on spatial data from geological records \citep{perez2013, Davis}. One of the transition detection methods that has been used widely for geological applications is the wavelet transform \citep{mallat1991}. In \citep{Arabjamaloei, Cooper} the wavelet transform technique has been used on petrophysical data (including gamma ray, sonic and resistivity records) to detect the geological transitions for identifying lithological boundaries (i.e. boundaries which seperate distinct rock units). \citep{perez2013} provides an introductory overview of using the wavelet transform for detecting significant transitions in geophysical data. A rectangular tessellation of the wavelet transform and a filtering of this tessellation to assign lithological domains in drill hole data is given by \citep{June}. In this research, we introduce the recurrence plot and quadrant scan methodologies for the purpose of detecting lithological boundaries. The recurrence plot and recurrence quantification analysis (RQA) methods have been successfully applied to detecting dynamical transitions in time series analysis of various systems, including financial analysis \citep{Gilmore}, engineering applications \citep{Elwakil}, chemistry \citep{Rustici} and applied physics \citep{Vretenar}. A comprehensive overview of these methods and their applications to complex systems (such as neuroscience, finance, geology, climate and others) is given by \citep{Marwan2, Marwan3}. The key motivation for proposing the recurrence plot and quadrant scan methodology is its applicability to multivariate data; it enables dimensionality reduction of the problem by integrating the multivariate geological data. This is an advantage over wavelet methods, which can only be applied to univariate data. The proposed method is computationally cheap and fast. We demonstrate that the method performs well in the presence of noise and that the technique does not depend on high data resolution to successfully identify the location of the boundaries.

\section{Data and study area} \label{sectiondata}
\begin{figure}
	\centering
	\includegraphics[width=\textwidth]{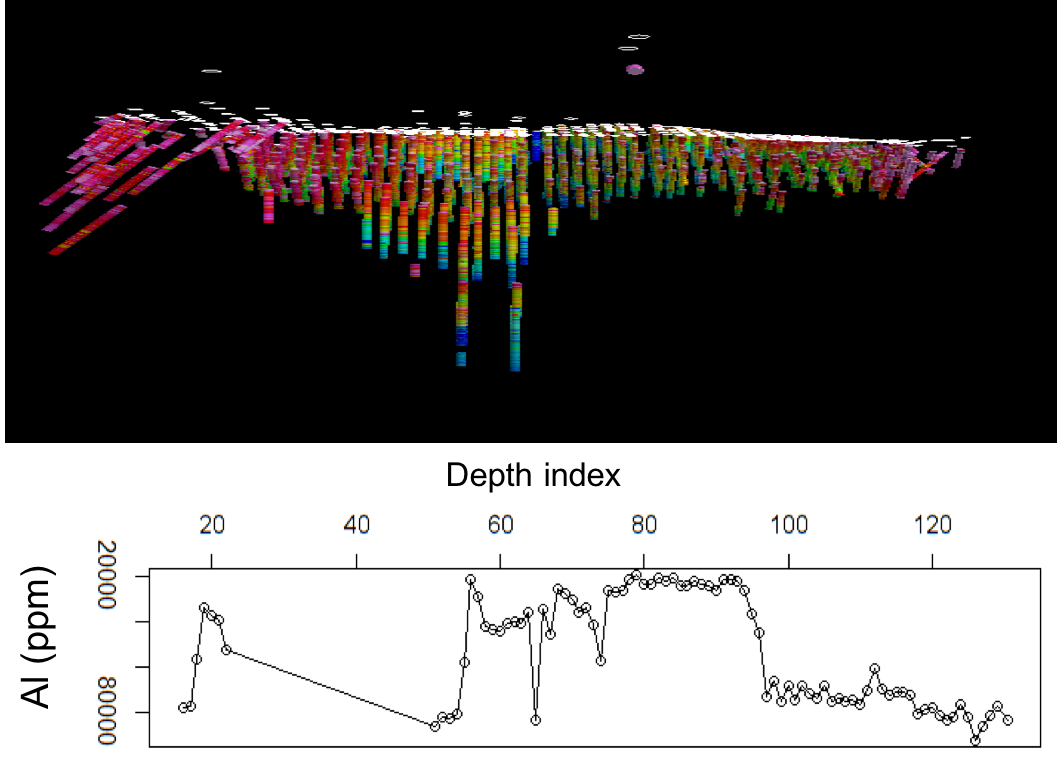}
        \caption{Geochemical drill hole data: (Top) Spatial distribution of Al content in 600 drill holes.(Bottom) Down hole plot of Al ppm for a single drill hole; the plot shows several sharp transitions in composition, representing lithological boundaries intersected by the drill hole.}
	\label{fig:drillholesfield}
\end{figure}

The proposed method has been tested on two different types of data. The first data set consists of multi-element geochemical data from a Ni-Cu-PGE deposit (i.e. nickel, copper and platinum group elements). The deposit lies in a sulphide-rich layer within a large mafic-ultramafic layered intrusion. The shape and geometric relationships (interpreted from drill hole data) between the intrusive layers, the Ni-Cu ore and the PGE ore are key to understanding the genesis of these ore bodies \citep{Barnes}. There are approximately 600 drill holes, comprising hundreds of samples. Each sample was taken over a 1 m interval of the core. Fig. \ref{fig:drillholesfield} displays 3D plots of the drill holes, viewed from the south and coloured by aluminium content  (Al ppm). The figure illustrates the layered nature of the intrusive where vertical changes occur over a shorter distance than lateral changes due to the presence of shallowly dipping layers. Fig. \ref{fig:drillholesfield} also contains a downhole plot of Al content from a single drill hole. This plot illustrates the abrupt changes in compositional values that indicate the presence of lithological boundaries. 

\begin{figure}
	\centering
	\includegraphics[width=\textwidth]{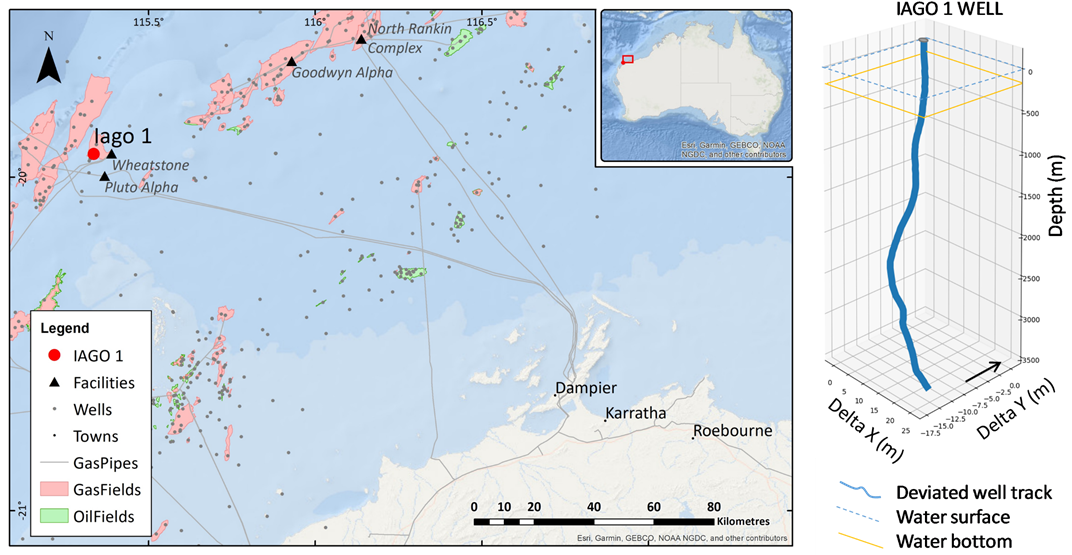}
        \caption{Location (left) and the 3D well track (right) of Iago 1 in North West Shelf, offshore Western Australia.}
	\label{fig:Iago1}
\end{figure}

The second data set is a set of petrophysical well logs from a deep offshore gas exploration well, Iago-1, from the Northwest Shelf of Australia (see Fig. \ref{fig:Iago1}).  The complete well dataset is found in the form of Log ASCII STANARD (.LAS) files at no cost from the publicly available WAPIMS (Western Australian Petroleum and Geothermal Information Management System) database. Six petrophysical measurements were chosen from the well that record changes in density, electrical resistivity, sonic velocity, natural radioactivity, mean atomic number and neutron porosity of the rocks penetrated by the well. These six logs are typically the most important and commonly acquired petrophysical measurements used in offshore oil and gas exploration wells. Similar to the mineral drill hole example, the well logs contain significant local changes in the petrophysical records due to lithological boundaries. However, these two data sets are different in scale and resolution. The drill hole data is sampled every metre and hundreds of points have been recorded for each hole. The maximum depth reached for any one hole is less than 400 m. The well log record spans $\sim$1800 m, commencing some distance below the seabed and the data sampling is every 15 cm. The well logs traverse a series of different sedimentary units and the gas-bearing zone is near the bottom of the well. For both data sets, the initial aim is to distinguish the domains representing lithological layers by identifying the most significant boundaries.

\section{Methodology: Density quadrant scan} \label{method}
\begin{figure}
        \centering
        \includegraphics[width=0.7\textwidth]{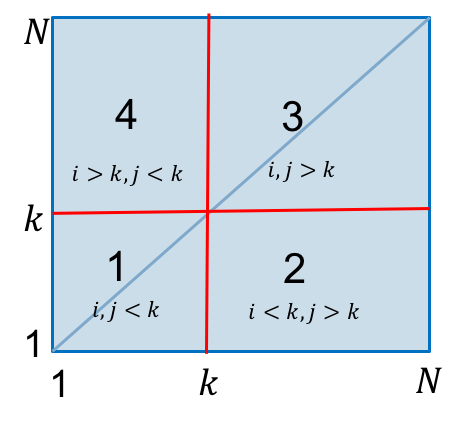}
        \caption{Schematic of the density quadrant scan method: at depth index $k$ we divide the recurrence matrix into four quadrants by two perpendicular axes. Then we compare the density of $1's$ in quadrants $1$ and $3$ to the density of $1's$ in all quadrants as presented in Eq. \ref{eq:densityQS}} 
        \label{fig:DQS}
    \end{figure}

We adopt and develop a time series technique to achieve the task of detecting geological boundaries from geochemical or petrophysical properties. Specifically, we use the well-known recurrence plot method \citep{Marwan2} with the addition of the quadrant scan procedure \citep{QS,QSDave,QSZaitouny}. The recurrence plot and the cross recurrence plot have been previously used once before on geological data in \citep{Marwan1} but for an entirely different purpose. The drill hole data is spatial data, hence we adjust the proposed methods to be applicable to spatial geological data sets. We consider depth as the independent variable, a proxy for time. The algorithm is based on the following essential steps:
\begin{itemize}
\item \textbf{Construct a Norm Matrix:} Let $x_{d_i} \in \mathcal{R}^m$ be the state of the system (the data values) at depth $d_i$ where $i=1,2,\dots,N$. We construct a $N \times N$ matrix $A$ with $a_{ij}=\|x_{d_i}-x_{d_j}\|$, here $\|.\|$ is the Euclidian norm; $m$ is the number of elements under consideration. If $m > 1$, i.e. multivariate analysis, then normalisation is required in order to allow all elements under consideration to contribute equally to the outcome. Normalisation scales each element in the range 0 and 1, with the sum of all values equal to 1. For a specific component of $x$, i.e. $x^j$, normalisation is given by\footnote{This simple normalisation scheme is applicable in our case where all quantities are positive. However, in other applications where negative quantities are considered, alternative normalisation may be required.}:
\begin{equation}
\bar{x}_{d_i}^j = \frac{x_{d_i}^j}{\sum_{k=1}^N x_{d_k}^j} \notag
\end{equation} 
\item \textbf{Construct a Recurrence Plot Matrix:} From the norm matrix $A$, we construct an $N \times N$ binary matrix $R$ with $r_{ij}=1$ iff $a_{ij} < \epsilon$ and $0$ otherwise. When applying the method on multiple variables or different materials, it is important to select an appropriate threshold, $\epsilon$, to reduce the effects of signal noise. In the literature, there are several paradigms for selecting the recurrence plot threshold; for example, closest $n$ neighbours is a common scheme \citep{Marwan2}. In this application, the recurrence plot threshold, $\epsilon$, is based on the distribution of the elements of $A$. To avoid outliers and so reduce the effect of noise, the threshold is chosen using:
\begin{equation} 
\epsilon = \alpha \times (\text{mean}(a_{ij})+3\times \text{std}(a_{ij})), \text{for all $i$ and $j$ where } 0 < \alpha < 1 \notag 
\end{equation}
In general, the selection of $\alpha$ is on a case--by--case basis as it depends on the scale required for the solution\citep{QSZaitouny}. In this application, the value of $\alpha$ is determined by the variation level of the layers to match the scale used by the geologists. Numerical trials on different samples of both data sets, indicated that $\alpha=0.25$ is an appropriate value for the geochemical data set and $\alpha=0.05$ is suitable for the petrophysical data. The different $\alpha$ values result from the different variation level in the geochemical data compared to the petrophysical data. In \citep{QSZaitouny}, we show that by modifying $\alpha$ the user can control the scale of the output that it is suitable for the application and can classify the boundaries into multiple scales.
\item \textbf{Extract Quadrant Scan Sequence:} From the recurrence plot matrix $R$,  a sequence $q(d_k)$ is constructed from the point density ratios between the points in the same quadrant (i.e. quadrants with $i,j<k$ or $i,j>k$) to the points in all quadrants of $R$. Fig. \ref{fig:DQS} is a schematic demonstration of how to estimate the quadrant scan:
\begin{equation}
q(d_k)=\frac{D_{1,3}}{D_{1,3}+D_{2,4}} \label{eq:densityQS}
\end{equation}
Where $D_{1,3}$ is the density of the points in quadrants 1 and 3 while $D_{2,4}$ is the density in quadrants 2 and 4. They are defined as follows:
\begin{eqnarray}
D_{1,3}&=&\frac{\sum\limits_{i,j<k}r_{ij}+\sum\limits_{i,j>k}r_{ij}}{(k-1)^2+(N-k)^2} \notag \\
D_{2,4}&=&\frac{\sum\limits_{i<k,j>k}r_{ij}+\sum\limits_{i>k,j<k}r_{ij}}{(k-1)\times(N-k)\times 2} \notag
\end{eqnarray}
 Maxima (peaks) of $q(d_k)$ correspond to transitions in the dynamics, which are interpreted to represent geological boundaries. The value of $q(d_k)$ is between 0 and 1.
\end{itemize}
An advantage of this methodology is that it is applicable to univariate as well as multivariate data. This method allows data integration by recasting multiple records into a single sequence where peaks indicate multivariate transitions. This compression and simplification of the multivariate data is particularly helpful when comparing and correlating lithological layers across multiple drill holes. 

\section{Applications and results}
In this section, the proposed method is applied to both geochemical and petrophysical data sets, which are measured at different resolutions as described in Sec. \ref{sectiondata}. We use these diverse applications to show the robustness and versatility of the method. Furthermore, we discuss the quality and consistency of the results for both univariate and multivariate inputs. We also discuss the reduction of data dimension by integrating the logs of different variables onto a single quadrant scan profile.

\subsection{Geochemical drill hole data}
The results of the analysis are demonstrated on a single drill hole from the geochemical data set. The initial analysis is run using 4 chemical elements (Al, Fe, Mg and Ca); these were selected as they are important elements for lithological (rock type) classification. The results of layer detection for individual elements is compared to the results for multivariate analysis in order to demonstrate the advantages of multivariate analysis over univariate analysis.
 \subsubsection{Single variable}
\begin{figure}
    \centering
    \begin{subfigure}[b]{0.345\textwidth}
        \centering
        \includegraphics[width=\textwidth]{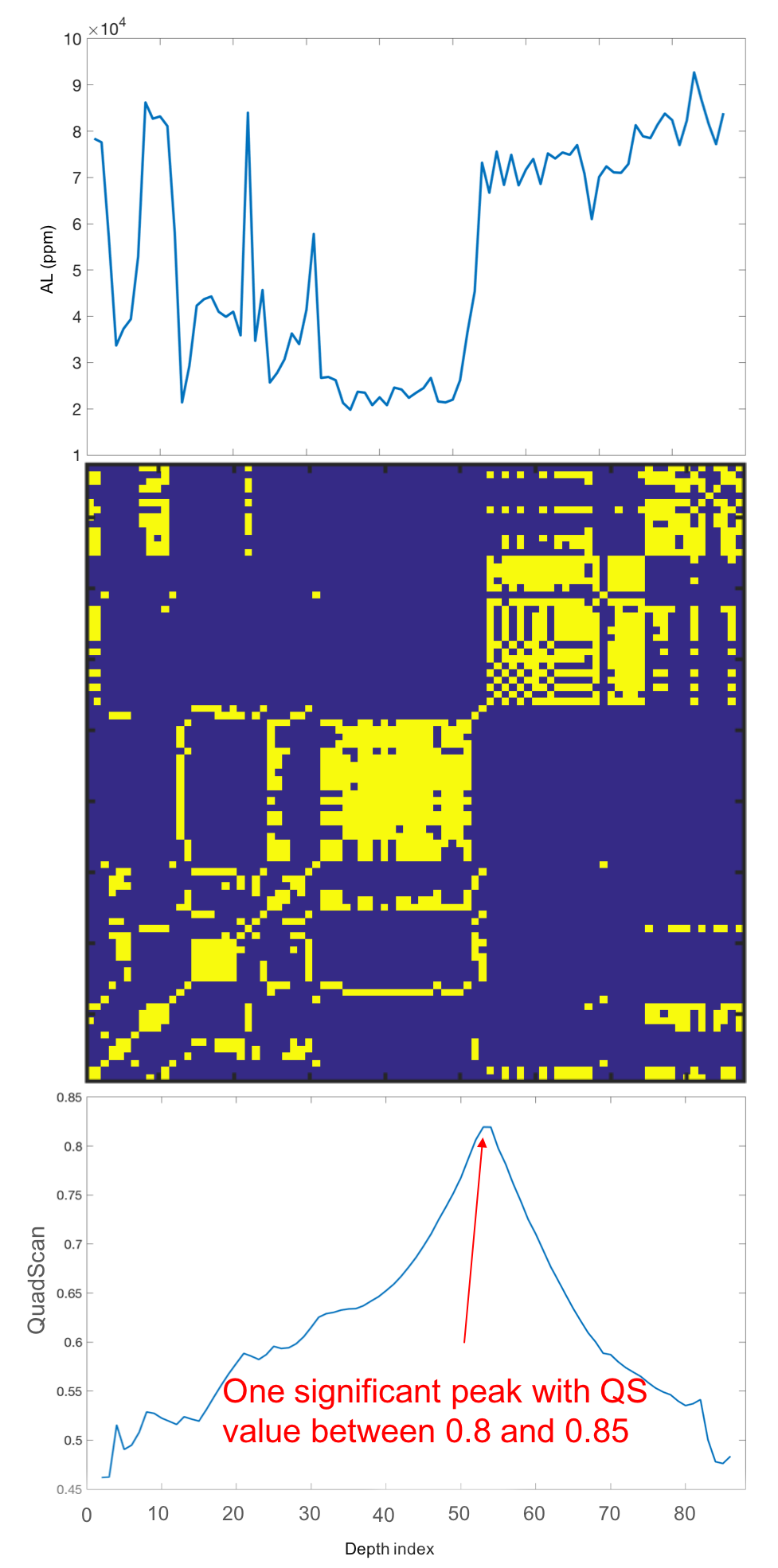}
        \caption{Al}
        \label{fig:Al}
    \end{subfigure}%
    ~ 
    \begin{subfigure}[b]{0.36\textwidth}
        \centering
        \includegraphics[width=\textwidth]{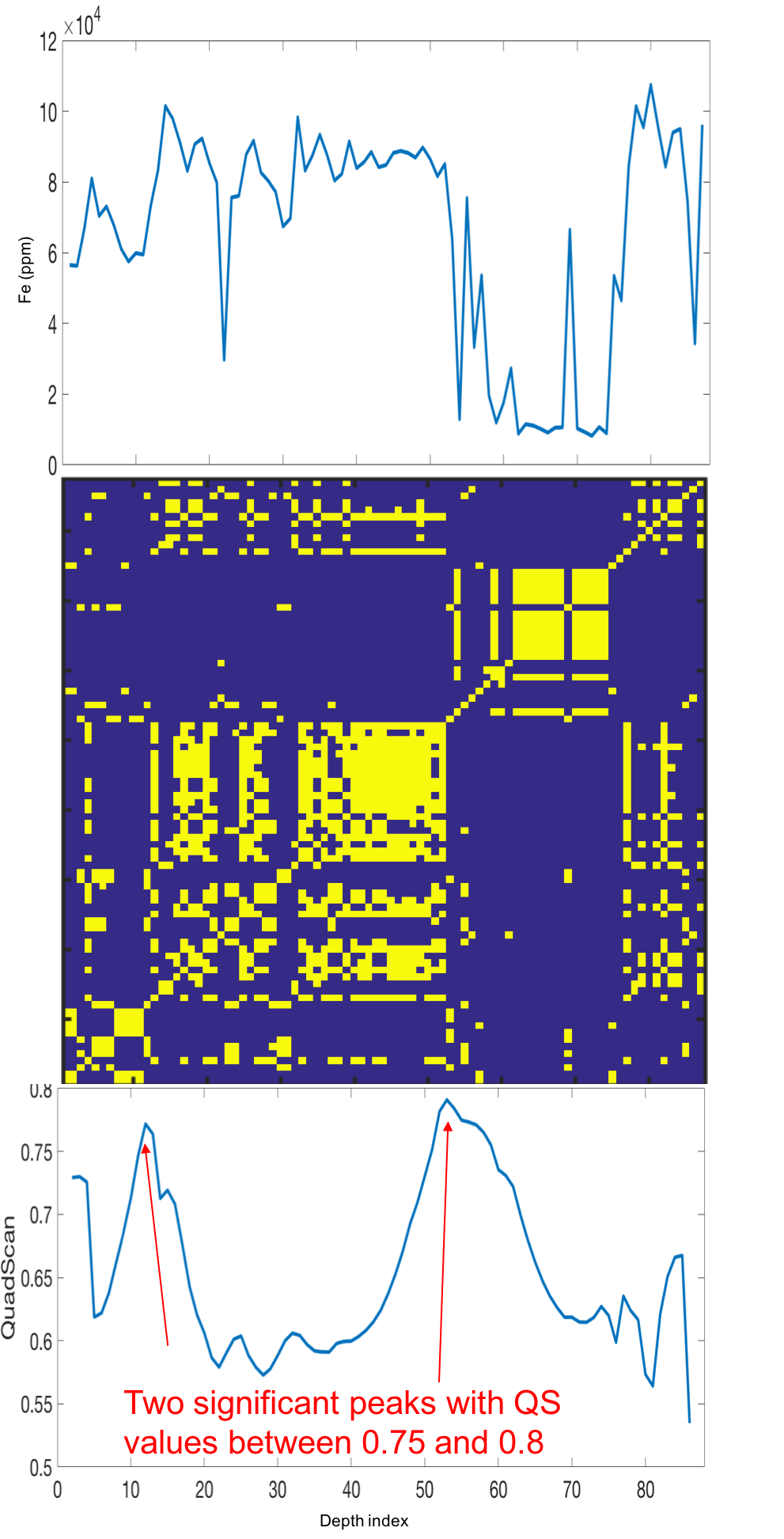}
        \caption{Fe}
        \label{fig:Fe}
    \end{subfigure}
    
     \begin{subfigure}[b]{0.33\textwidth}
        \centering
        \includegraphics[width=\textwidth]{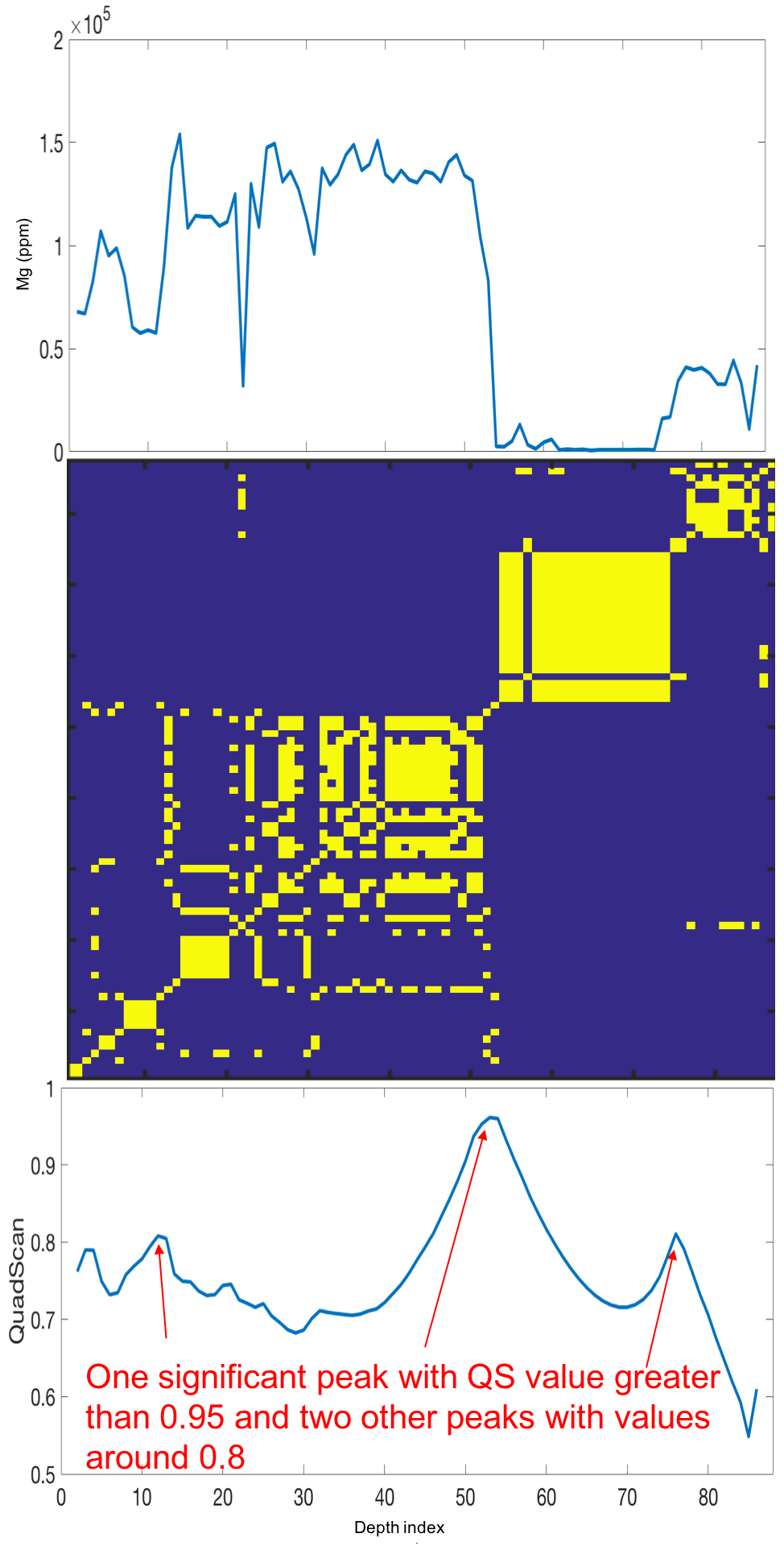}
        \caption{Mg}
        \label{fig:Mg}
    \end{subfigure}%
    ~ 
    \begin{subfigure}[b]{0.33\textwidth}
        \centering
        \includegraphics[width=\textwidth]{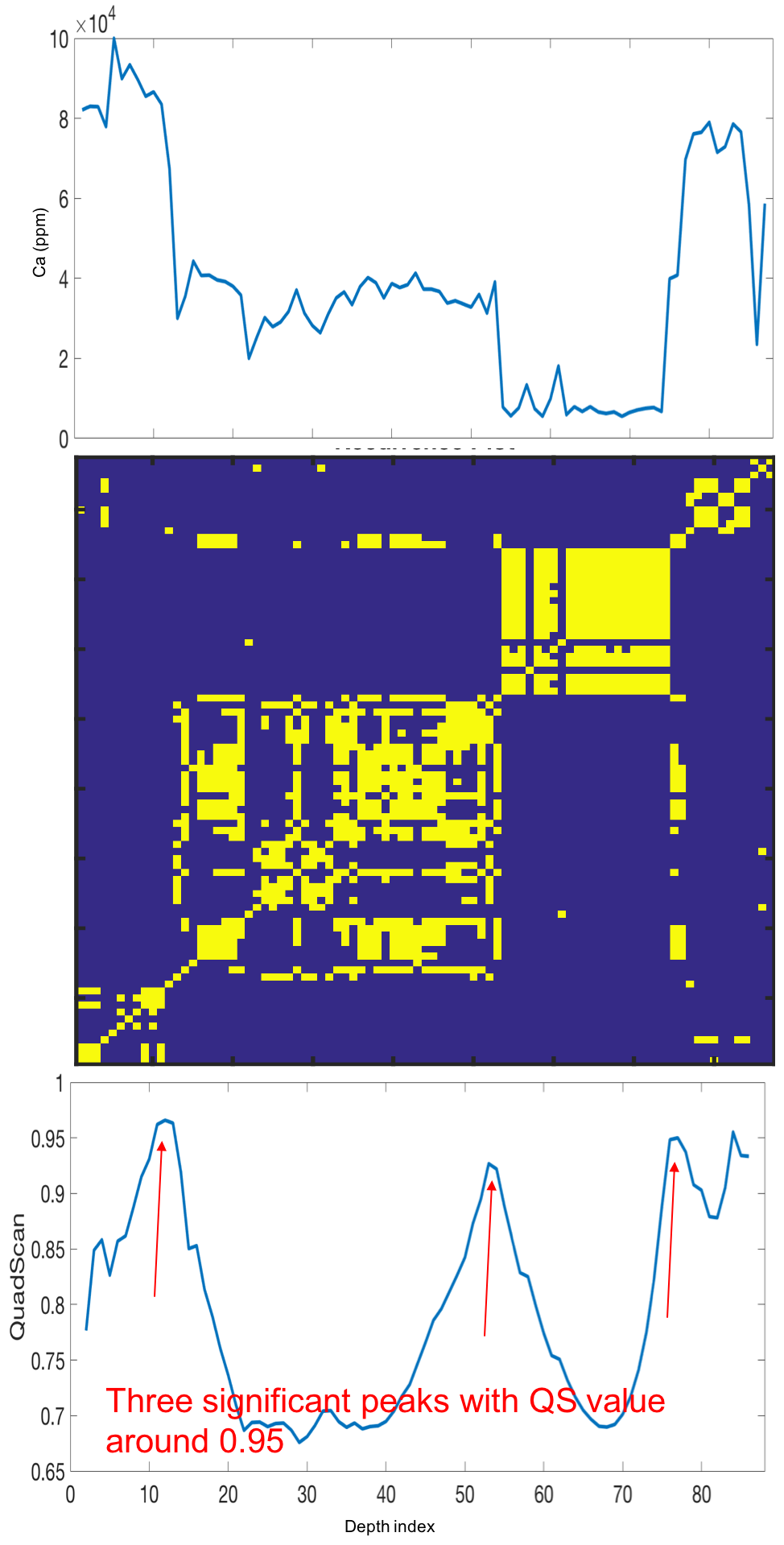}
        \caption{Ca}
        \label{fig:Ca}
    \end{subfigure}
    \caption{Results of implementation on single deposits from same drill hole. Top plots are the data records, middle plots are the recurrence plots and bottom plots are the corresponding quadrant scan sequences. Significant peaks are indicated by arrows.}
    \label{fig:single}
\end{figure}

 The results of the recurrence matrix and the quadrant scan of the four elements (Al, Fe, Mg and Ca) are illustrated in Fig. \ref{fig:single}. Al has a single significant transition point, which is common to all four elements, it occurs at sample number 52 (i.e. at a depth of 95 m). Additional transition points occur in Fe, Mg and Ca, which do not necessarily occur at common locations. For example,  the transition at sample 75 in Mg and Ca was not detectable in Al and Fe (Figs. \ref{fig:Al}, \ref{fig:Fe}). The solution to this problem is to combine the normalised elements into a single analysis.
  
\subsubsection{Multiple variables}
\begin{figure}
    \centering
    \begin{subfigure}[b]{0.31\textwidth}
        \centering
        \includegraphics[width=\textwidth]{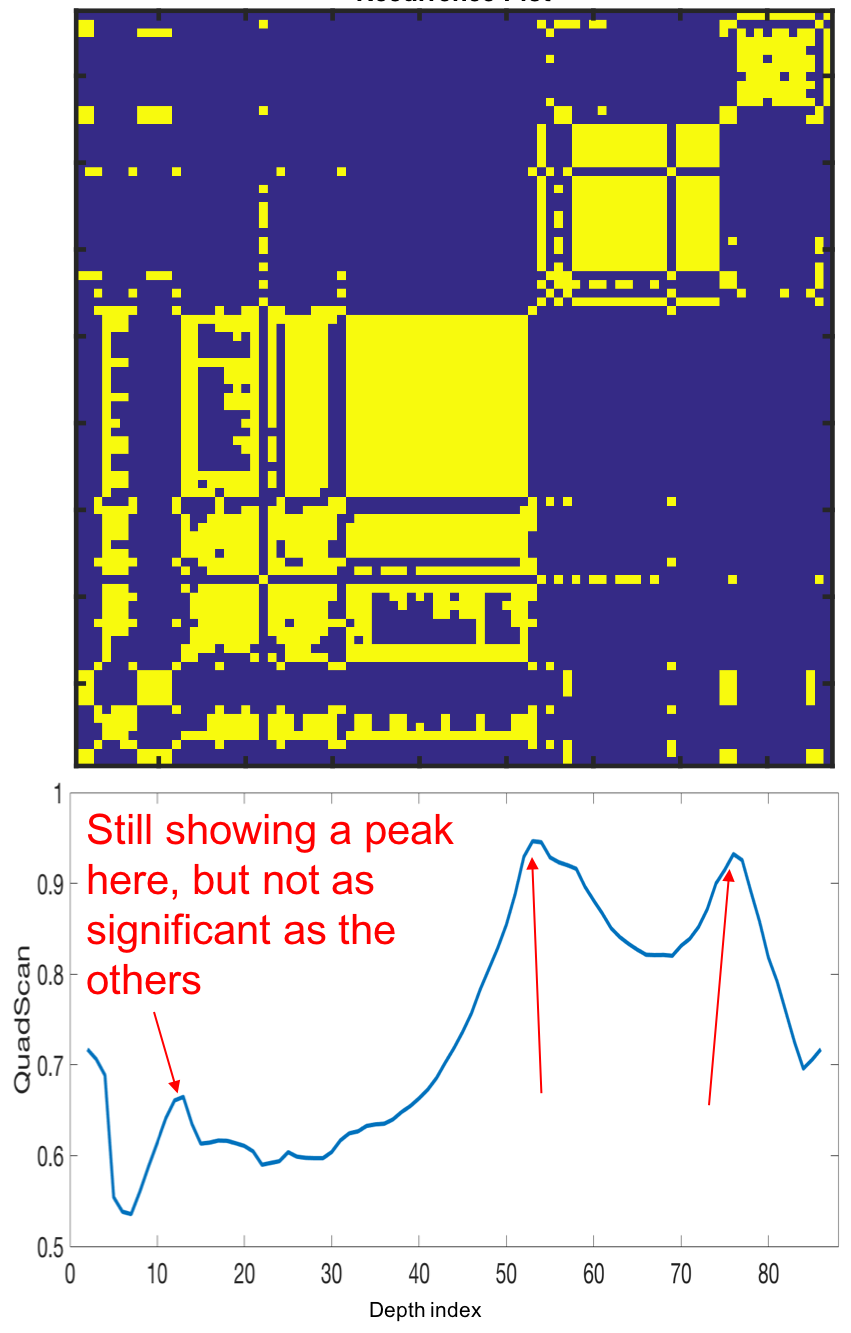}
        \caption{Al+Fe}
        \label{fig:AlFe}
    \end{subfigure}%
    ~ 
    \begin{subfigure}[b]{0.32\textwidth}
        \centering
        \includegraphics[width=\textwidth]{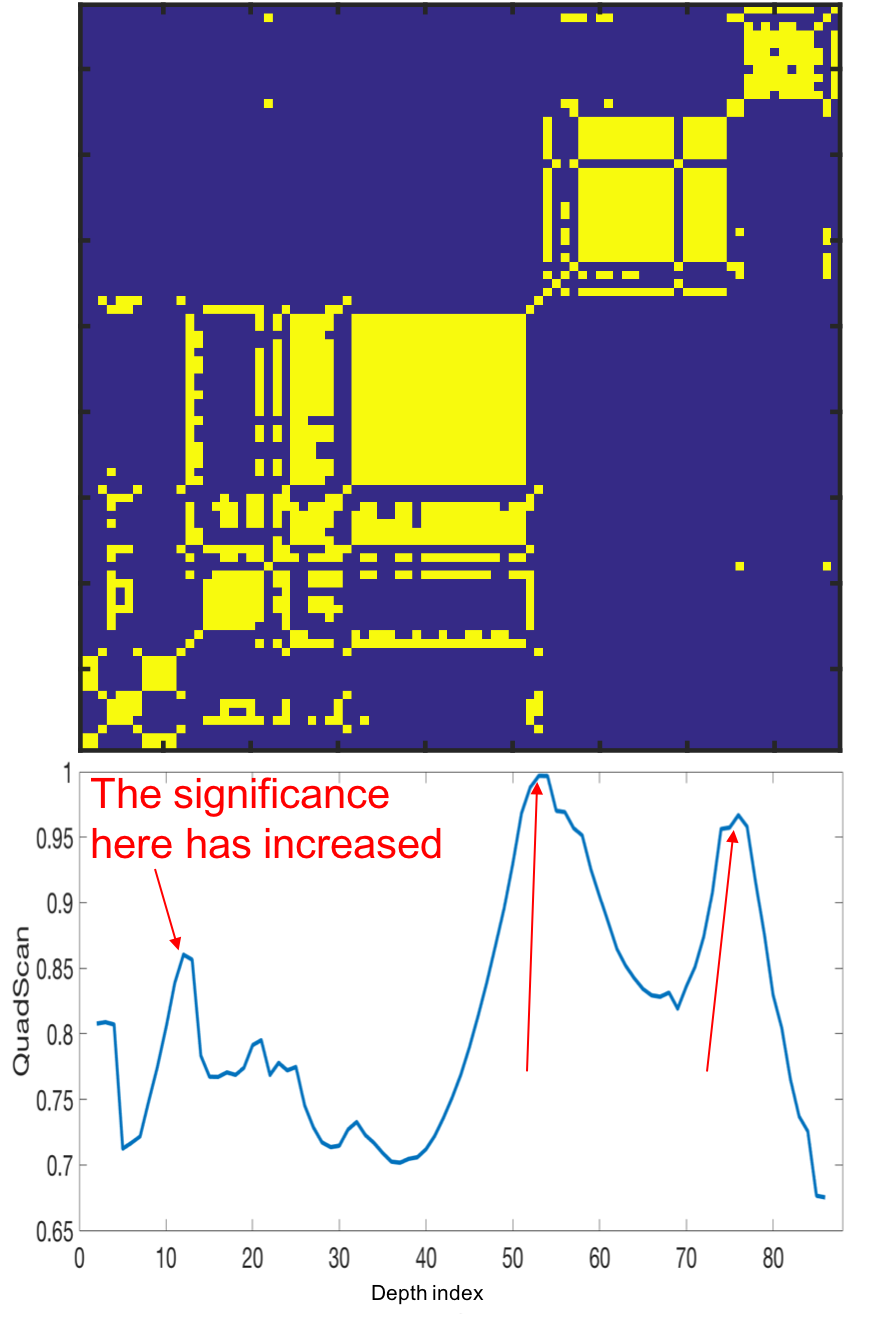}
        \caption{Al+Fe+Mg}
        \label{fig:AlFeMg}
    \end{subfigure}
    ~ 
    \begin{subfigure}[b]{0.315\textwidth}
        \centering
        \includegraphics[width=\textwidth]{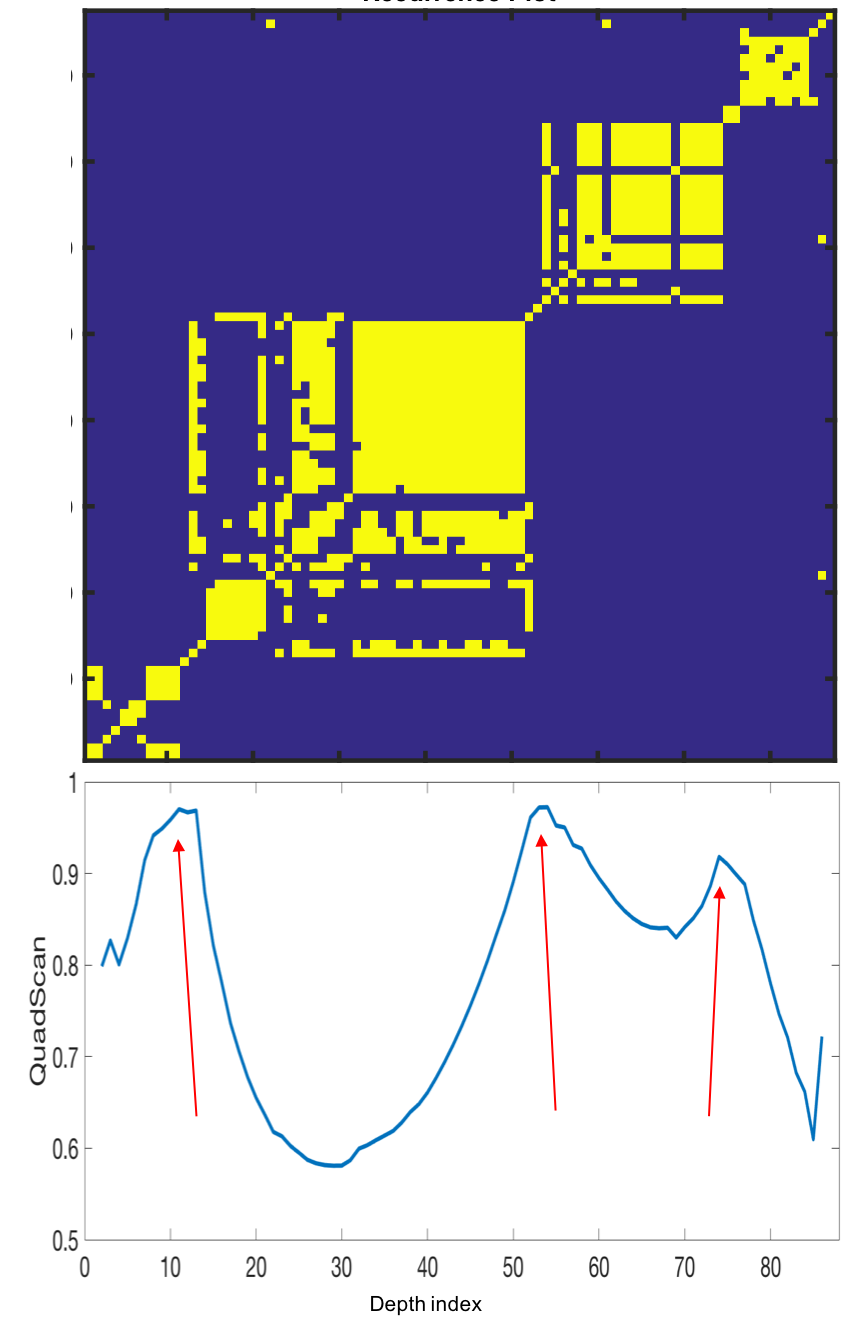}
        \caption{Al+Fe+Mg+Ca}
        \label{fig:AlFeMgCa}
    \end{subfigure}
    \caption{Results of implementation on multi--elements from same drill hole. Top plots are recurrence plots and bottom plots are corresponding quadrant scans. In (a) Al and Fe content are used as an input, in (b) Al, Fe and Mg are the input and in (c) all four elements are used as the input (i.e. Al, Fe, Mg, Ca). The peaks that indicate significant transitions are indicated by arrows.}
 \label{fig:multi}
\end{figure}

The result of combining different numbers of elements is illustrated in Fig. \ref{fig:multi}. The progressive inclusion of more information, in the form of element data, results in reduced noise in the recurrence plots and improved detection of all 3 significant transition points by the quadrant scan.
The multivariate results confirm the presence of three transition points that divide the drill hole into four distinct geological layers. 

\subsection{Petrophysical well log data}
\begin{figure}
    \centering
    \begin{subfigure}[b]{0.75\textwidth}
        \centering
        \includegraphics[width=\textwidth]{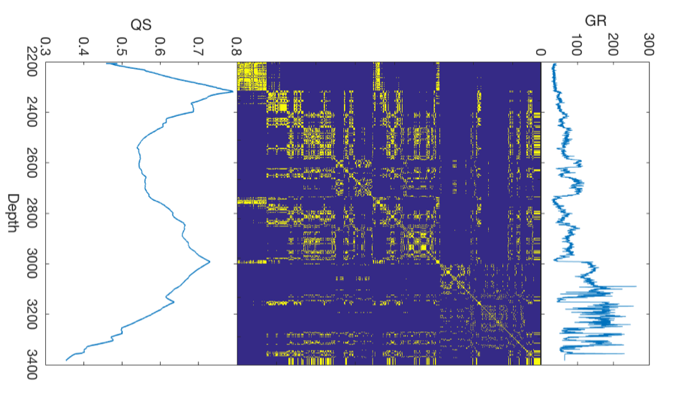}
        \caption{Gamma Ray}
        \label{fig:IagoGR}
    \end{subfigure} 
    \\
    \begin{subfigure}[b]{0.75\textwidth}
        \centering
        \includegraphics[width=\textwidth]{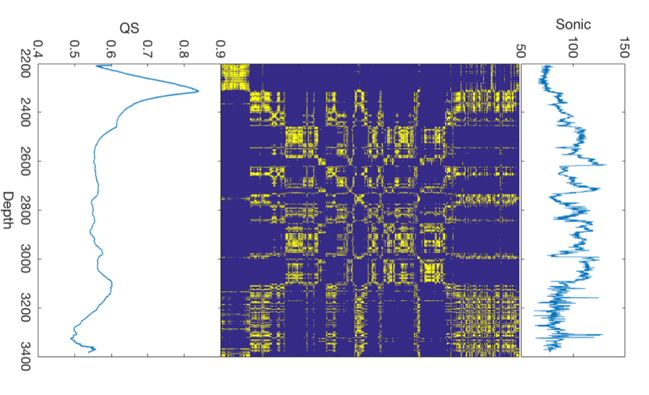}
        \caption{Sonic}
        \label{fig:IagoSonic}
    \end{subfigure}
    \\
    \begin{subfigure}[b]{0.75\textwidth}
        \centering
        \includegraphics[width=\textwidth]{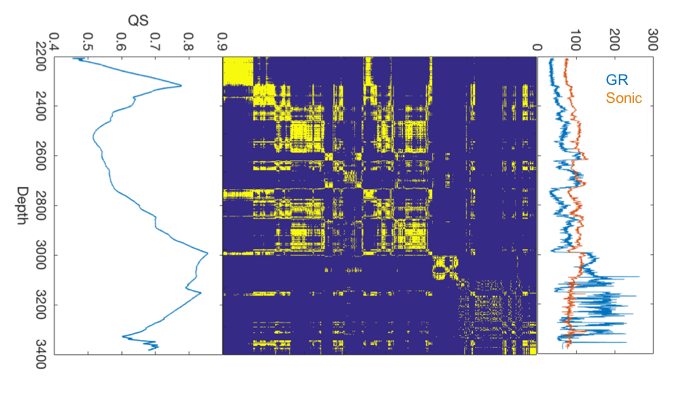}
        \caption{Gamma Ray and Sonic}
        \label{fig:IagoGRSonic}
    \end{subfigure}
    \caption{Results of implementation on petrophysical well-logs. Tops are the petrophysical logs, middle are the corresponding Recurrence Plots and bottoms are corresponding Quadrant Scans. In (a) Gamma Ray logs are used as an input, in (b) Sonic logs are the input and in (c) Gamma Ray and Sonic logs are combined and used as a multivariate input.}
 \label{fig:Iago}
\end{figure}

\begin{figure}
        \centering
        \includegraphics[width=1\textwidth]{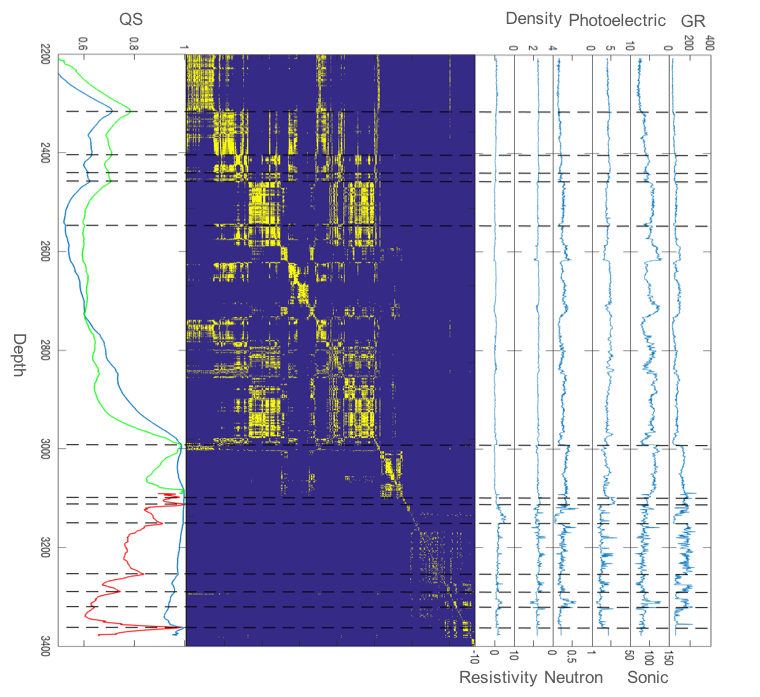}
        \caption{Gamma Ray, Sonic, Photoelectric, Neutron, Density and Resistivity logs are combined and used as a multivariate input. The blue quadrant scan resulted from the entire logs, the green and red resulted from running the quadrant scan scheme on two intervals as indicated.}
        \label{fig:IagoGRSPNDR}
    \end{figure}

The method is applied to petrophysical data from well logs (Iago-1). The petrophysical data set has a small sampling interval (15 cm) and consequently a very large number of measurement points (more than 7500 samples). The method is run on univariate and multivariate petrophysical logs. Fig. \ref{fig:Iago} shows the results from individual analysis of Gamma Ray logs (panel a) and Sonic logs (panel b). The results demonstrate common significant peaks where very sharp transitions occur. However, the quadrant scan profiles show other lower peaks indicative of less sharp transitions. In Fig. \ref{fig:IagoGRSonic} Gamma Ray and Sonic logs are combined to improve our detection results, the resulting recurrence plot and quadrant scan indicate more consistent detection of the layers. Finally, in Fig. \ref{fig:IagoGRSPNDR} additional petrophysical properties have been included, namely photoelectric, neutron porosity, density and resistivity. All detected peaks (high and low), where potential transitions occur, have been highlighted, the results indicate that adding more petrophysical features into the analysis yields more consistent detection of the transitions relating to boundaries of petrophysical domains.

The well log data clearly has two regions of distinctive variance, with lower variance above 3100 m, and higher variance below this depth; the deeper, higher variance section corresponds to the oil and gas reservoir (Fig. \ref{fig:IagoGRSPNDR}). The analysis was subsequently run separately on these two regions to test if this improves transition detection. The results of separate analysis of the deeper region (in red) have been enhanced and the transitions have been emphasised. However, along the interval between depth 2600 m to 3000 m the detected peaks are still underestimated in the quadrant scan profile. Therefore, in the following section we propose an improvement to deal with these challenges and enhance the quadrant scan outcomes.

\section{Improving the quadrant scan: weight quadrant scan}
Although the methodology has demonstrated successful detection of geological boundaries, some transitions have been underestimated as the repetition in the signal decreases the value of some peaks (see Fig. \ref{fig:IagoGRSPNDR} the middle interval of depth between 2600 to 3000 m). This happens because the density quadrant scan (Eq. \ref{eq:densityQS}) counts all the past and future points equally. This is reasonable and important in time series analysis. However, in the case of transition detection from spatial records, the method underestimates some transitions which have local, rather than global, importance. Therefore, in this section we propose a development of the quadrant scan in order to improve the detection and overcome the effect of the repetition in the spatial signals. For each depth index $k$ we assign distance weights to the points in each quadrant instead of counting the density of the points. This method increases the impact of nearby points and decreases the effect of relatively distant points. To avoid the computational cost for assigning a weight for each point for every depth $d_k$, we assign a matrix weight for each quadrant instead, as follows:

For depth $d_k$ where $k=2 \dots N-1$ we define the weighting vectors:
\begin{eqnarray}
V1(l_1)&=&\frac{1}{2}(1-\tanh(((k-l_1)-m_1)/m_2)) \notag \\
V2(l_2)&=&\frac{1}{2}(1-\tanh((l_2-m_1)/m_2)) \notag
\end{eqnarray}
where $l_1=1\dots k-1$ and $l_2=1:N-k$. $m_1$ and $m_2 \in \mathbb{R}$ are parameters to tune the weighting scheme. The modified $\tanh$ weighting function provides an initial flat segment at 1, which at a point determined by the parameters $m_1$ and $m_2$, decreases gradually to 0. This method is based on a similar weighting scheme in \citep{flock}. The result is that points in the near neighbourhood contribute significantly to the final value, while very distant points have no contribution. The tuning parameters are not very sensitive and they determine (1) the size of the neighbourhood where the weight equals to 1 and (2) the slope of the gradual decrease to 0. For the examples tested, values of $m_1=200$ and $m_2=50$ provided a good result. These values assign a weight equal to 1 for approximately 100 closest points either side of the point under consideration; beyond this the weights decrease gradually to reach 0 for approximately 600 points either side.\\
The following matrices are used to calculate the quadrant scan:
\begin{eqnarray}
W1&=&V1^T \times V1 \notag \\
W2&=&V1^T\times V2 \notag \\
W3&=&V2^T \times V2 \notag \\
W4&=&V2^T \times V1 \notag \\
R1&=&(r_{ij}) \text{ for } i,j<k \notag \\
R2&=&(r_{ij}) \text{ for } i<k,j>k \notag \\
R3&=&(r_{ij}) \text{ for } i,j>k \notag \\
R4&=&(r_{ij}) \text{ for } i>k,j<k \notag 
\end{eqnarray}
The first four matrices are the weighting matrices assigned for each quadrant (resulting from an appropriate multiplication of the weighting vectors and their transposed vectors). The latter four matrices represent the corresponding quadrants from the recurrence matrices. Using these matrices, the quadrant scan is calculated as follows:
\begin{equation}
q(d_k)=\frac{\sum R1\circ \frac{W1}{w1_{k-1,k-1}}+\sum R3\circ (\frac{W3}{w3_{1,1}})}{\sum R1\circ  \frac{W1}{w1_{k-1,k-1}}+\sum R3\circ \frac{W3}{w3_{1,1}}+\sum R2\circ \frac{W2}{w2_{k-1,1}}+\sum R4\circ \frac{W4}{w4_{1,k-1}}}
\end{equation} 
where $\circ$ denotes element--wise multiplication\footnote{We divide the weighting matrices by the largest entry in each matrix to normalise the weights between 0 and 1}.
\begin{figure}
        \centering
        \includegraphics[width=0.65\textwidth]{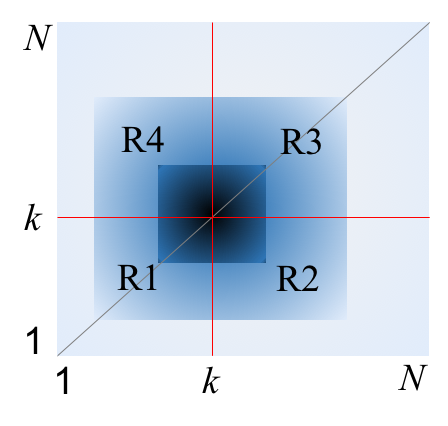}
        \caption{Schematic demonstration of the weight quadrant scan method. Darker boxes include points with higher weights, closest points are assigned with a weight equal to 1 then the weights decrease gradually to reach 0 at the farthest points.}
        \label{fig:WQS}
    \end{figure}

\begin{figure}
        \centering
        \includegraphics[width=1\textwidth]{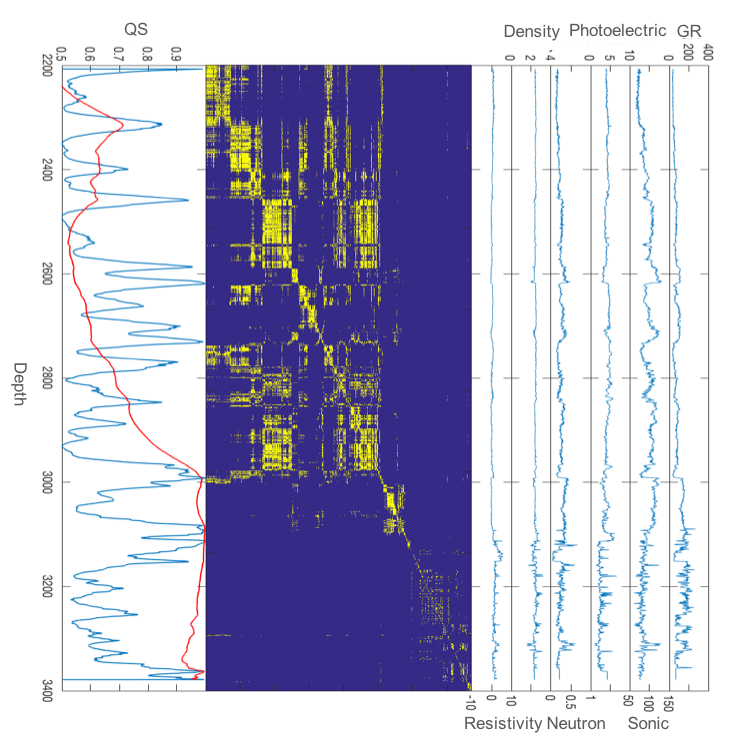}
        \caption{Demonstration of the results using weight quadrant scan (in blue) compared with the density quadrant scan (in red).}
        \label{fig:Iagoweighted}
    \end{figure}

Fig. \ref{fig:Iagoweighted} compares the results of both schemes. It is clear that the weighted quadrant scan improves the results significantly by increasing the prominence of the previously underestimated transitions, resulting in improved lithological layer separation.

\section{Discussion} \label{Advantages}
\subsection{Comparison with geological logs}
The results of the proposed methodology for multivariate transition detection are compared against a geologist's manual interpretation of the well log. The manual interpretation is biostratigraphy, i.e. the boundaries represent changes in microfossil assemblages within the well core samples. It is important to note that the input data for the manual interpretation (i.e. microfossils) is different to the input data for the transition analysis, wireline logging of petrophysical properties. However, there should be a close correspondence between the two methods as microfossil assemblages should correspond to changes in environment, which would also be reflected in changes in sedimentary processes and hence changes in lithology in the sedimentary sequence.

\begin{figure}
        \centering
        \includegraphics[width=1\textwidth]{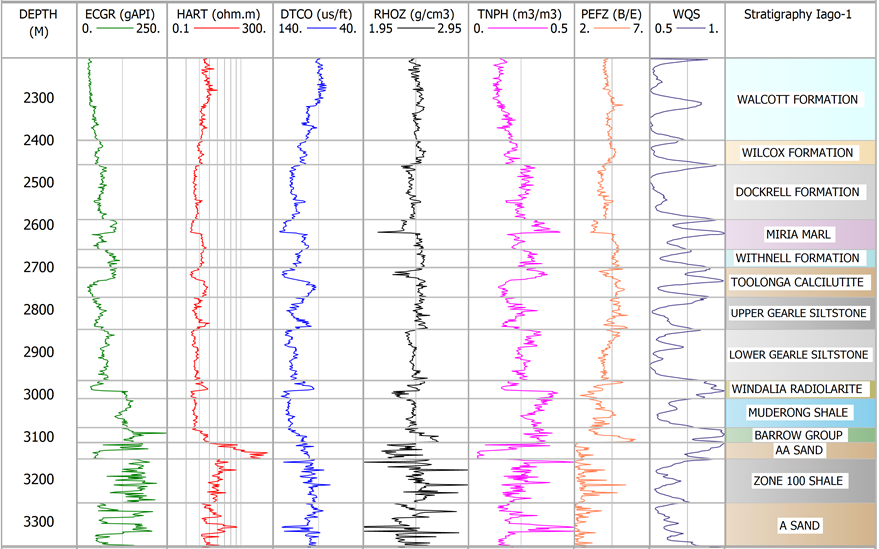}
        \caption{Comparison of the transition detection results between the weight quadrant scan (WQS) and the biostratigraphic boundaries. The quadrant scan profile is plotted as the blue line on the right. The horizontal gray lines on the data profiles represent the biostratigraphic layers.}
        \label{fig:QSvsgeology}
    \end{figure}

Comparison between the biostratigraphy and the transition analysis is illustrated in Fig. \ref{fig:QSvsgeology}. The results indicate that the transition analysis is capable of detecting most of the changes recorded in the biostratigraphy.

The quadrant scan results identify additional transitions, which correspond to a change in the petrophysical rock type, without a corresponding change in the microfossil assemblage. These additional transitions correspond to subtle changes in rock type, for example at 2890 m (in the Lower Gearle Siltstone), or major changes in physical properties that do not have a corresponding change in the microfossil assemblage (e.g. 2615 m in the Miria Marl). Some boundaries detected by the method cannot be related to a simple change in physical properties above and below, and therefore may point out locations in the well worthwhile to examine in more detail, e.g. within the shaley sequence at $\sim$3200 m (Zone 100 Shale).

It is known that the fluid content changes from water to gas, at approximately 3135 m. However, no boundary is detected by the quadrant scan, probably because the gas content is variable and not persistent with depth. The base of the gas zone at 3150 m is detected as a major boundary as it coincides with a change in rock type from sandstone containing gas, to shale below. On the other hand, Hill et al. \citep{June} developed a method of multiscale domaining that has been applied to well log data, and we have investigated how recurrence plot methods perform by comparison. The multiscale domaining approach is good for detecting local contrasts in properties (e.g. gas sands are identified as distinct), whereas the recurrence plot methods we introduce here more easily finds similar rock types that are spatially separated. Indeed, the two methods are complimentary in many respects and can be used together to bring a complete classification of rock types and identify their boundaries.

\subsection{Advantages}
The transition detection method has a number of advantages over existing methods. In comparison to manual geological logging it is fast and can produce consistent, repeatable results. In comparison to univariate wavelet transformation methods, it has been demonstrated to work well with multivariate (geochemical or petrophysical) data, which helps to reduce noise as well as reduce dimensionality. The method demonstrates consistency in transition detection under addition of more variables, with existing peaks being retained and additional transition peaks being resolved from the added variables. The fact that this method integrates the phase space of a single bore-hole (i.e. multiple geochemical or petrophysical measures) into a quadrant scan sequence as well as its ability to characterise multiple--scale boundaries \citep{QSZaitouny} allow us, as a future work, to use the resulting sequences in further investigations to correlate the boundaries and rock types across different holes (wells). 

Changes in the specific physical properties of density and sonic velocity are responsible for the patterns of reflections seen in seismic surveys across a sedimentary basin. These are interpreted to understand the structure of the basin (sedimentary packages, subsided or uplifted areas, folds and faults). Therefore, an objective method for identifying the boundaries in well data and comparing these with seismic reflections can be used iteratively to search for significant boundaries across the basin and predict where these may be encountered in new wells.

Moving beyond seismic imaging the physical properties depend not only on the rock minerals and porosity, but also on the content of fluids such as oil and gas. Quantitative interpretation of seismic data using the amplitude and phase of reflectivity patterns, or by mathematical inversions of the seismic waveforms, is used by petroleum explorationists to directly detect, hydrocarbon fluids, or at least improve the probability of success when drilling new wells. The methods described here can help to standardize and speed up through automation, the process of defining the main rock packages, and within these packages the distribution of distinct and potentially hydrocarbon-bearing rock types. Boundaries at scales seen in the well data occur at high resolutions and at larger scales seen by the seismic data, which has a spatial resolution typically of only 10-20 metres. Therefore, adaptive matching of boundaries from the recurrence plot method can be used to tune into features seen at seismic scale in two or three dimensions across the basin.

\section{Conclusion}
A multivariate time-series technique has been developed to detect lithological layers from petrophysical and geochemical profiles from drill hole or well log data. The method is unsupervised so does not require any prior geological information. The quadrant scan method successfully detects geological boundaries from both drill hole and well log data sets. The method allows variables to be combined to achieve more consistent results compared to laborious traditional methods involving manual geological logging and petrophysical analysis. The method has potential for general application as it is useful for any depth-attributed numerical data, including geochemical and petrophysical data, and it has been demonstrated to work successfully on data with very different sampling resolutions and depth spans.

The modified weight quadrant scan overcomes common challenges in practical applications; it allows successful detection of large scale as well as small scale transitions, without producing dense small transitions where changes are geologically insignificant. This reduces the negative impact of different levels of variation and noise possibly present in the data. In addition, varying the parameter $\alpha$ allows multi-scale detection, that is, the boundaries can be characterised based on their strengths which is crucial when multiple holes are considered. Finally, this versatile method is efficient, accurate and computationally cheap and fast. Although in general a point--wise weighting scheme based on distances for an $N\times N$ matrix is time consuming, the alternative weighting scheme that we propose using the vectors $V1$ and $V2$ and the corresponding weighting sub-matrices $W1,W2,W3,W4$ overcomes this computational issue and significantly reduces the calculation time. 

Although the method demonstrates several advantages as mentioned above, it is important to emphasise the non--trivial modifications required of recurrence plot and quadrant scan methods. Specifically, since we are dealing with spatial data, we need to take into account the non--embedded and non--causal nature of the data, which are not a time series. However, using multivariate spatial data provides the same kind of advantages, and facilitates the use of recurrence plot methods for the efficient analysis of drill hole data and petrophysical well logs.

These single--hole analyses open the door for future investigations to complete the big picture by considering multiple holes.  We can utilise the dimensionality reduction and multi-scale characterisation abilities together with the cross recurrence plots and adjusted Quadrant Scan to analyse the cross-correlation between different holes.

\section*{Acknowledgement}AZ and MS acknowledge the support of the Australian Research Council through Discovery Grant DP 180100718. 

\section*{Code availability}The MATLAB code is available in the following public GitHub repository\\
 https://github.com/AyhamZaitouny/Boundaries-Detection-Weight-Quadrant-Scan-


\bibliographystyle{unsrt}
\bibliography{refs}

\end{document}